\documentstyle[epsf]{mn}

\newif\ifAMStwofonts

\ifoldfss
   \ifCUPmtlplainloaded \else
     \NewTextAlphabet{textbfit} {cmbxti10} {}
     \NewTextAlphabet{textbfss} {cmssbx10} {}
     \NewMathAlphabet{mathbfit} {cmbxti10} {} 
     \NewMathAlphabet{mathbfss} {cmssbx10} {} 
   \fi
   \ifAMStwofonts
     \ifCUPmtlplainloaded \else
       \NewSymbolFont{upmath} {eurm10}
       \NewSymbolFont{AMSa} {msam10}
       \NewMathSymbol{\upi}     {0}{upmath}{19}
       \NewMathSymbol{\umu}     {0}{upmath}{16}
       \NewMathSymbol{\upartial}{0}{upmath}{40}
       \NewMathSymbol{\leqslant}{3}{AMSa}{36}
       \NewMathSymbol{\geqslant}{3}{AMSa}{3E}

     \fi
   \fi
\fi 

\ifnfssone
   \newmathalphabet{\mathit}
   \addtoversion{normal}{\mathit}{cmr}{m}{it}
   \addtoversion{bold}{\mathit}{cmr}{bx}{it}
   \newmathalphabet{\mathbfit} 
   \addtoversion{normal}{\mathbfit}{cmr}{bx}{it}
   \addtoversion{bold}{\mathbfit}{cmr}{bx}{it}
   \newmathalphabet{\mathbfss} 
   \addtoversion{normal}{\mathbfss}{cmss}{bx}{n}
   \addtoversion{bold}{\mathbfss}{cmss}{bx}{n}
   \ifAMStwofonts
     \ifCUPmtlplainloaded \else
       \UseAMStwoboldmath
       \makeatletter
       \new@mathgroup\upmath@group
       \define@mathgroup\mv@normal\upmath@group{eur}{m}{n}
       \define@mathgroup\mv@bold\upmath@group{eur}{b}{n}
       \edef\UPM{\hexnumber\upmath@group}
       \new@mathgroup\amsa@group
       \define@mathgroup\mv@normal\amsa@group{msa}{m}{n}
       \define@mathgroup\mv@bold\amsa@group{msa}{m}{n}
       \edef\AMSa{\hexnumber\amsa@group}
       \makeatother
       \mathchardef\upi="0\UPM19
       \mathchardef\umu="0\UPM16
       \mathchardef\upartial="0\UPM40
       \mathchardef\leqslant="3\AMSa36
       \mathchardef\geqslant="3\AMSa3E
     \fi
   \fi
\fi 

\ifnfsstwo
   \DeclareMathAlphabet{\mathbfit}{OT1}{cmr}{bx}{it}
   \SetMathAlphabet\mathbfit{bold}{OT1}{cmr}{bx}{it}
   \DeclareMathAlphabet{\mathbfss}{OT1}{cmss}{bx}{n}
   \SetMathAlphabet\mathbfss{bold}{OT1}{cmss}{bx}{n}
   \ifAMStwofonts
     \ifCUPmtlplainloaded \else
       \DeclareSymbolFont{UPM}{U}{eur}{m}{n}
       \SetSymbolFont{UPM}{bold}{U}{eur}{b}{n}
       \DeclareSymbolFont{AMSa}{U}{msa}{m}{n}
       \DeclareMathSymbol{\upi}{0}{UPM}{"19}
       \DeclareMathSymbol{\umu}{0}{UPM}{"16}
       \DeclareMathSymbol{\upartial}{0}{UPM}{"40}
       \DeclareMathSymbol{\leqslant}{3}{AMSa}{"36}
       \DeclareMathSymbol{\geqslant}{3}{AMSa}{"3E}
     \fi
   \fi
\fi 

\ifCUPmtlplainloaded \else
   \ifAMStwofonts \else 
     \def\upi{\pi}
     \def\umu{\mu}
     \def\upartial{\partial}
   \fi
\fi

\title[A CTT star in the $\eta$ Cha cluster]{
ECHA J0843.3--7905: Discovery of an `old' classical
T Tauri star in the $\eta$ Chamaeleontis cluster}
\author[W. A. Lawson et al.]
{Warrick A. Lawson$^{1}$, Lisa A. Crause$^{2}$,
Eric E. Mamajek$^{3}$ and Eric D. Feigelson$^{4}$\\
$^{1}$School of Physics, University College UNSW, Australian
Defence Force Academy, Canberra ACT 2600, Australia\\
$^{2}$Department of Astronomy, University of Cape Town,
Private Bag, Rondebosch 7700, South Africa\\
$^{3}$Steward Observatory, University of Arizona, 933 N Cherry
Avenue, Tuscon AZ 85721, USA\\
$^{4}$Department of Astronomy and Astrophysics, Pennsylvania State
University, University Park PA 16802, USA\\
{\rm E-mail: wal@ph.adfa.edu.au (WAL);
lcrause@artemisia.ast.uct.ac.za (LAC);
eem@as.arizona.edu (EEM); edf@astro.psu.edu (EDF)}}
\date{Accepted 2001 October 24. Received 2001 October 18;
       in original form 2001 July 19}

\pagerange{\pageref{firstpage}--\pageref{lastpage}}
\pubyear{2001}

\begin{document}

\maketitle

\label{firstpage}

\begin{abstract}
A limited-area survey of the $\eta$ Chamaeleontis
cluster has identified 2 new late-type members.  The
more significant of these is ECHA J0843.3--7905 (= {\it IRAS\,}
F08450--7854), a slowly-rotating ($P = 12$ d) M2 classical T
Tauri (CTT) star with a spectrum dominated by Balmer emission.
At a distance of 97 pc and cluster age of $\approx 9$ Myr, the
star is a nearby rare example of an `old' CTT star and promises
to be a rewarding laboratory for the study of disk structure
and evolution in pre-main sequence (PMS) stars.  The other new
member is the M4 weak-lined T Tauri (WTT) star ECHA J0841.5--7853,
which is the lowest mass ($M \approx 0.2$ M$_{\odot}$) primary
known in the cluster.
\end{abstract}


\section[]{Introduction}

The $\eta$ Chamaeleontis star cluster is a compact grouping
of PMS stars at a distance of 97 pc (Mamajek, Lawson \&
Feigelson 1999, 2000).  The cluster is one of the closest open
clusters, and one of the nearest groups of PMS stars known.
Analysis of Hertzsprung-Russell (HR) diagrams constructed
using contemporary evolutionary models indicate an age of
$\approx 9$ Myr (Lawson \& Feigelson 2001), similar to other
nearby groups of PMS stars such as the TW Hya association
(Webb et al. 1999).  The age of the cluster is intermediate
between young PMS stars (ages of $<$ a few Myr) still associated
with their parent molecular clouds and older post-T Tauri
(ages of $>$ a few $\times 10$ Myr) populations, and is thus
at an interesting age for the study of science issues concerning
early stellar evolution, e.g., angular momentum evolution, disk
dissipation and planet formation, and magnetic activity evolution.

Mamajek et al. (1999) listed a population of 13 primaries; 12
of these are X-ray emitting stars detected during a {\it ROSAT\,}
High-Resolution Imager (HRI) pointing (the RECX stars) and the
13th is the X-ray-quiet A star HD 75505.  These 13 objects
consist of 3 early-type systems ($\eta$ Cha, RS Cha AB and HD
75505) and 10 WTT stars.

Completing the census of cluster members is important for studies
examining the properties of a coeval population of PMS stars.
Consideration of the cluster initial mass function (IMF) suggests
that the stellar population may be $2-4 \times$ the known number
of primaries (Mamajek et al.  2000).  In this paper we report the
discovery of 2 new late-type cluster members residing within
the {\it ROSAT\,} HRI region, following a search for stars with
photometric chacteristics similar to the known members.  In the
following sections we detail our search and analysis methods.

\section[]{Observations and Data Reduction}

\subsection{Photometric selection of candidate members}

Compared to most field stars of similar colour (or temperature),
members of the $\eta$ Cha cluster are elevated in magnitude (or
luminosity) in the colour-mag (or HR) diagram, owing to a combination
of youth, proximity and low interstellar reddening.  [Westin (1985)
found $E$($b-y$) = --0.004 for $\eta$ Cha, indicating that reddening
is unimportant for the cluster.]  Cluster members appear
to be highly coeval (to $\pm\, 1-2$ Myr depending upon the adopted
evolutionary model; see Lawson \& Feigelson 2001), with most of
the late-type RECX stars forming a near-linear sequence in the
colour-magnitude diagram.  Several RECX stars appeared elevated in
brightness above the sequence due to suspected binarity, which
has since been confirmed in 2 cases by K\"ohler (2001).

These characteristics can be used to select candidate cluster members
with similar photometric properties to the known cluster stars.
Spectroscopic study can then confirm signs of stellar youth, in
particular enhanced lithium.  Thus new members of the cluster
can be found irrespective of their X-ray properties and without
prior knowledge of their distances or proper motions.  However,
such a sample might suffer contamination from background giants
and foreground stars.  Mamajek et al. (2000; see their fig. 4)
showed a colour-magnitude diagram for $>$ 21,000 stars within a
1$^{\rm o}$ radius of the cluster based upon United States Naval
Observatory (USNO) A2.0 photographic photometry (Monet et al. 1998).
The RECX stars were elevated above the vast majority of stars,
but a broad giant branch overlapped the region of the
diagram occupied by the K-type RECX stars.

Lawson et al. (2001) used the 1-m telescope and 1k $\times$ 1k
SITe charge-coupled device (CCD) at the South African Astronomical
Observatory (SAAO) to obtain multi-epoch differential $V$-band
observations and Cousins {\it VRI\,} photometry of the late-type
members of the cluster.  In addition to these data, fields
neighbouring those containing the known members were observed
to search for new members.

{\it VRI\,} frames were obtained with exposure times ranging
from $1-300$ s to encompass the magnitude range of
the late-type members, and to obtain photometry to fainter
limits.  The deepest frames obtained useful (few percent
accuracy) photometry of stars of $V \approx 18$, or
stars of spectral type M5 and mass $M \approx 0.1$M$_{\odot}$
at the distance and age of the cluster according to the models
of Siess, Dufour \& Forestini (2000).  These data were calibrated
against equatorial and southern photometric standard stars.
The SITe CCD has a field of view of 26 arcmin$^{2}$ at the $f$/16
Cassegrain focus of the 1-m telescope.  23 fields were observed,
many over-lapping, for a total area of $\approx 500$ arcmin$^{2}$
or $\approx$ 40\% of the extent of the cluster as defined by
Mamajek et al. (2000).  Stars in each field were compared to the
sequence of RECX stars in the ($V-I$) versus $V$ colour-magnitude
diagram.  Candidates were selected as those objects with $V$
mags that fell, for their ($V-I$) colour, within the range
$-1.0 < V < 1.0$ mag of the sequence of RECX stars (Fig. 1).
This criterion accounted for the possibility of a greater
spread of ages within the cluster than is apparent in the RECX
stars, highly reddened stars or binary stars with elevated $V$
mags, and non-linearities in the colour-magnitude relationship
for the cluster (Lawson \& Feigelson 2001).  Of $\approx 2000$
stars measured, only 6 met the criterion (Table 1).  Most stars
in the fields were $> 2$ mag fainter than the sequence; in
particular the 3 M-type candidates were the only stars with
($V-I$) $> 2$, and with $V$ magnitudes that differed by
$< 2.5$ mag from the extrapolated linear sequence.

The differential $V$-band data were obtained with a 30 s
exposure time, with $\approx 40$ observations of each
field obtained during 1999 and $\approx 25$ during 2000.
The methods used to reduce and analyse the calibrated and
differential photometric data are discussed by Lawson et al.
(2001).

\begin{figure}
\begin{center}
\epsfxsize=8.4cm
\epsffile{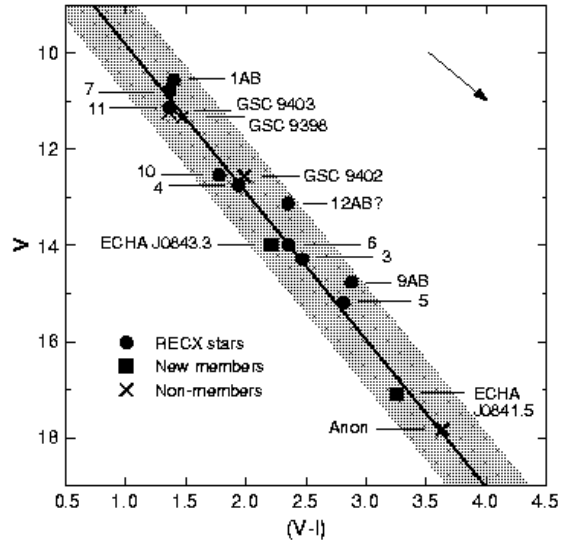}
\caption{Colour-magnitude diagram for the known and candidate
late-type members of the $\eta$ Cha cluster.  The bold line is
the extrapolated linear sequence of the RECX stars.  The grey
window highlights the region of the diagram searched for candidate
members.  The arrow is the reddening vector $A_{V}$/$E$($V-I$) =
1.93 (Rieke \& Lebofsky 1985).  The RECX stars are denoted by
number, with known (RECX 1 and 9) and suspected (RECX 12) binaries
indicated.  See Table 1 for the full identification of the
photometric candidates.}
\end{center}
\end{figure}

\begin{table*}
\centering
\caption{J2000 positions, Cousins {\it VRI\,} photometry, and
spectroscopic features of the candidate members of the $\eta$
Cha cluster.  Spectral types have an estimated uncertainty of
$\pm$ 1 subtype.  The final column lists confirmed new members.}
\begin{tabular}{@{}lcccccrrcc@{}}
\hline
      & $\alpha_{2000}$ & $\delta_{2000}$ & & & &
      \multicolumn{1}{c}{H$\alpha$ {\it EW}} &
      \multicolumn{1}{c}{Li I {\it EW}} & Sp. & \\
Star & (h~m~s) & (d~m~s) & $V$ & ($V-R$) & ($V-I$) &
\multicolumn{1}{c}{(\AA\,)} & \multicolumn{1}{c}{(\AA\,)} &
type & Comments \\
\hline
GSC 9402\_1003 & 08~36~40.7 & --78~54~58 & 12.55 & 0.96 & 1.98 &
0.9~~ & $< 0.1$~~ & M2  & \\
GSC 9398\_0099 & 08~37~51.4 & --78~44~37 & 11.33 & 0.76 & 1.46 &
1.1~~ & 0.0~~ & K5  & \\
USNO Anon 1 & 08~41~30.6 & --78~53~07 & 17.07 & 1.52 & 3.25 &
$-12$~~~~ & 0.9~~ & M4  & ECHA J0841.5--7853 \\
USNO Anon 2 & 08~43~18.4 & --79~05~21
			              & 13.97 & 0.99 & 2.20 &
$-110$~~~~ & 0.6~~ & M2  & ECHA J0843.3--7905 \\
Anonymous      & 08~48~34.9 & --78~53~52 & 17.81 & 1.69 & 3.63 &
$-3.5$~~  & $< 0.1$~~ & M5  & \\
GSC 9403\_0831 & 08~51~23.2 & --79~05~23 & 11.21 & 0.72 & 1.35 &
1.0~~ & 0.0~~ & K4 & \\
\hline
\end{tabular}
\end{table*}

\subsection{Spectroscopic confirmation}

Optical spectroscopy of the 6 candidates was obtained on
2000 April 21 and 22 using the 2.3-m telescope and dual-beam
spectrograph at Mount Stromlo and Siding Springs Observatories
(MSSSO).  In the red beam, the 1200R (1200 line\,mm$^{-1}$)
grating gave a 2-pixel resolution of 1.1 \AA\, with coverage
from $\lambda\lambda 6200-7160$ \AA.  In the blue beam, the 600B
(600 line\,mm$^{-1}$) grating gave a 2-pixel resolution of 2.2 \AA\,
with coverage from $\lambda\lambda 3840-5400$ \AA.  Exposure times
ranged from 720 s to 3000 s, and yielded continuum signal-to-noise
(S/N) ratios of $15-130$ in the red and, for the 4 brightest stars,
$10-40$ in the blue.  The spectra were calibrated using dome-flats,
bias frames and Fe-Ar arc frames, making use of standard library
routines such as {\tt ccdproc} within {\tt IRAF}.

Analysis of the spectra showed 2 of these stars were active,
lithium-rich late-type objects.  One of these stars (listed as
USNO Anon 1 in Table 1) has only a USNO-A2.0 catalogue entry.
The other star (listed as USNO Anon 2 in Table 1) appears to
be the optical counterpart of the {\it IRAS\,} Faint Source
Catalogue ({\it IRAS\,} FSC; Moshir et al. 1989) object {\it IRAS\,}
F08450--7854.  The {\it IRAS\,} FSC position differs by 5 arcsec
from the USNO position, with a $11 \times 2$ arcsec error ellipse.
The {\it IRAS\,} source had been mistakenly associated with the
nearby early-type cluster member RS Cha AB.  Neither star was
detected by {\it ROSAT\,} HRI indicating X-ray luminosities log
$L_{\rm X} < 28.5$ erg\,s$^{-1}$ (Mamajek et al. 2000).  These
2 stars (henceforth ECHA J0841.5--7853 and ECHA J0843.3--7905,
respectively) are assigned an $\eta$ Cha cluster designation
making use of J2000 coordinates.  Identification fields for
both stars, derived from the Second Palomar Observatory Sky
Survey (POSS-II), are shown in Fig. 2.  The other photometric
candidates are discussed in Section 3.3.

\begin{figure}
\begin{center}
\epsfxsize=8.0cm
\epsffile{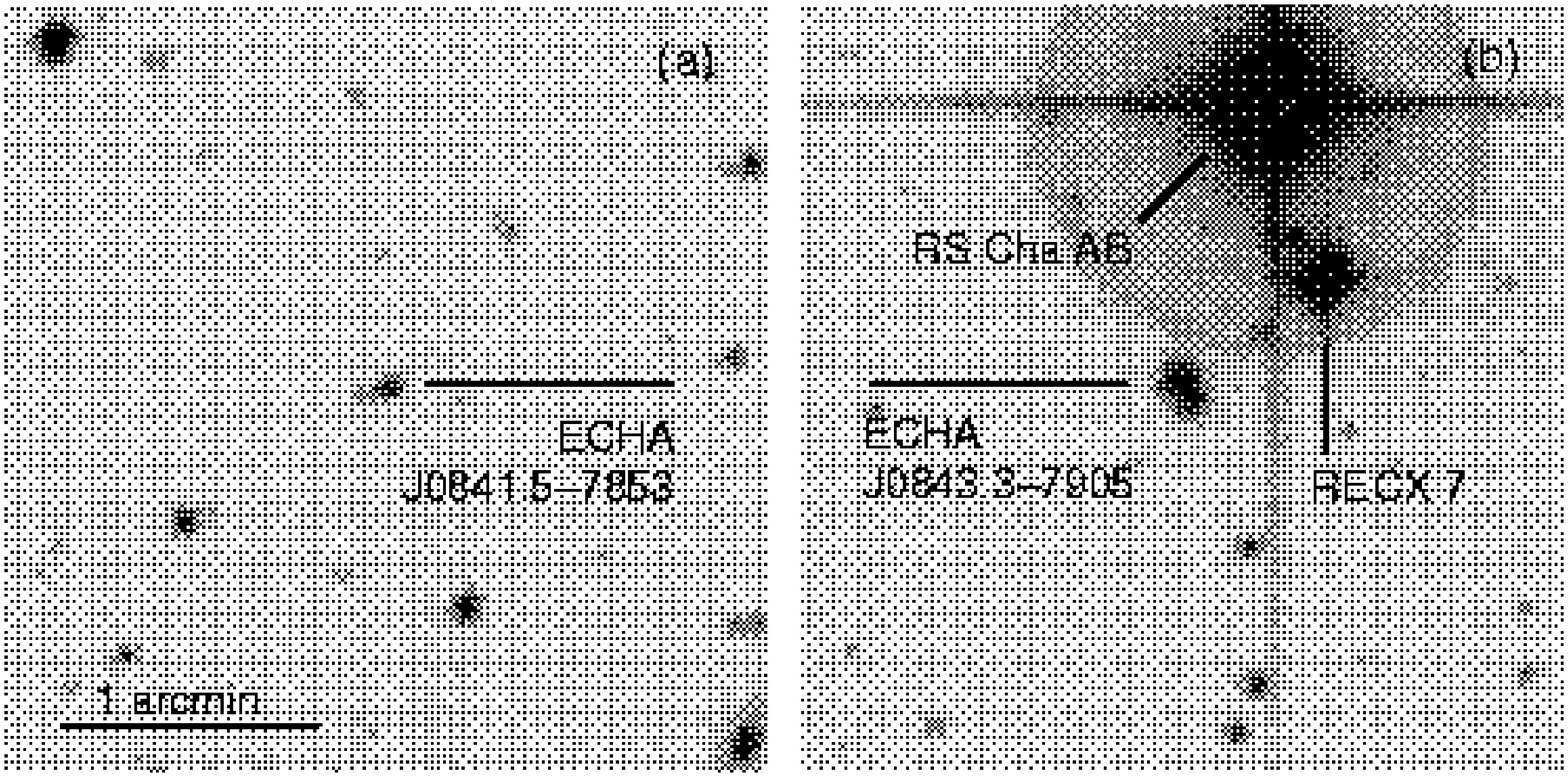}
\caption{POSS-II finder charts (width = 3 arcmin) centred on the
positions of (a) ECHA J0841.5--7853 and (b) ECHA J0843.3--7905.
In (b), cluster members RS Cha and RECX 7 are
identified.  The optical `companions' 8 arcsec E of ECHA
J0841.5--7853 and 6 arcsec SW of ECHA J0843.3--7905 are
unrelated field stars.}
\end{center}
\end{figure}

\section{Discussion}

\subsection{Spectroscopy of the new cluster members}

\subsubsection{ECHA J0841.5--7853}

Spectroscopy of ECHA J0841.5--7853 (Fig. 3) shows the star to be
of spectral type $\approx$ M4.  The level of optical activity
is typical for a late-M WTT star, with the H$\alpha$ equivalent
width ({\it EW\,}) = $-12$ \AA.  $\lambda 6707$ Li I absorption
is strong ({\it EW\,} = 0.9 \AA) and the spectrum also shows
weak $\lambda 6300$ [OI] emission.  The blue spectrum of the star
(not shown) shows weak H$\beta$ and H$\gamma$ emission; however
the poor S/N ratio of the spectrum prevented reliable measurement
of these lines.  At $V = 17.1$ and an inferred mass
$M \approx 0.2$ M$_{\odot}$ from Siess et al. (2000) tracks, the
star is the optically faintest and lowest mass primary known in
the cluster.

\subsubsection{ECHA J0843.3--7905}

The more significant of the 2 new members is ECHA J0843.3--7905.
The star is clearly a CTT star with an optical spectrum dominated
by strong Balmer and Ca II emission (Fig. 4).  The H$\alpha\/EW
= -110$ \AA, and the H$\beta\/EW = -26$ \AA.  Also present in
emission in the blue spectrum (Fig. 4a) is $\lambda 4063$ Fe I,
$\lambda 4471$ He I and all three of the chromospheric lines of
the Fe II (42) multiplet ($\lambda 4923$, 5018 and 5169).  The
red spectrum (Fig. 4b) shows strong $\lambda 6300$, 6363 [OI]
emission, and emission lines of [N II], He I (the $\lambda 6678$
transition is shown; also the $\lambda 7065$ transition is present)
and [S II].  $\lambda 6707$ Li I is present in absorption with
{\it EW\,} = 0.6 \AA.

Consideration of the spectrum of the star yields a spectral type
of M2$-$M3.  The colour of the star ($V-I \approx 2.2$) yields a
similar result ($\approx$ M2).  This classification is tentative
given the likely high level of optical veiling in the spectrum.
The star also has a high degree of optical variability, which we
discuss below.  From the colour-magnitude diagram and Siess et al.
(2000) models, the inferred mass of the star is $M = 0.3-0.4$
M$_{\odot}$.

\begin{figure}
\begin{center}
\epsfxsize=8.4cm
\epsffile{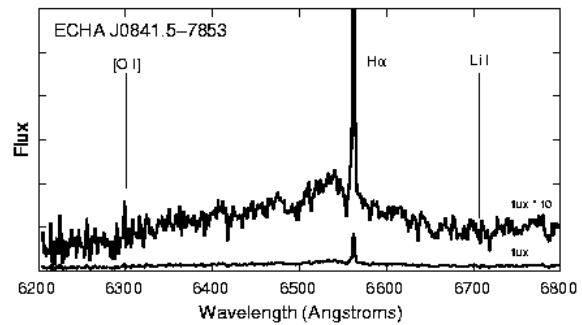}
\caption{Spectrum of the WTT star ECHA J0841.5--7853 near
H$\alpha$, shown here at reduced resolution.  Key emission
and absorption lines in the spectrum are identified.  The
spectrum is also shown scaled by a factor of 10 to highlight
weak emission and absorption features.}
\end{center}
\end{figure}

\begin{figure}
\begin{center}
\epsfxsize=8.4cm
\epsffile{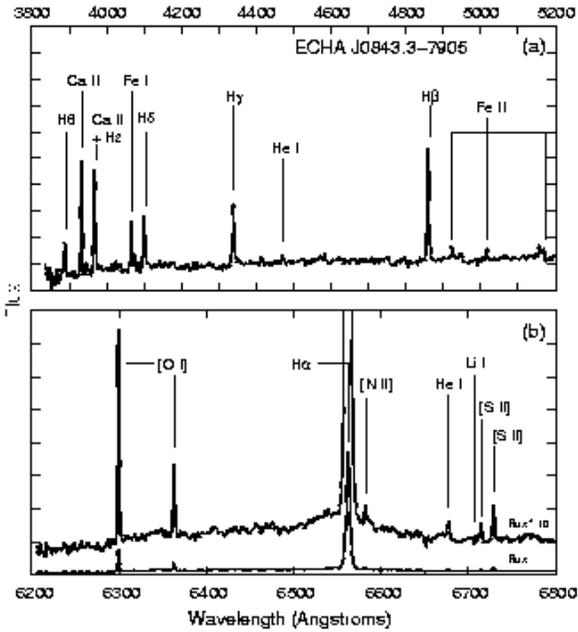}
\caption{(a) Blue and (b) red spectrum of the CTT
star ECHA J0843.3--7905, shown here at reduced resolution.
Key emission and absorption lines are identified.  In (b),
the spectrum is also shown scaled by a factor of 10 to
highlight weak emission features and the presence of
$\lambda 6707$ Li I.}
\end{center}
\end{figure}

\subsection{Multi-epoch {\it V\,} observations}

Multi-epoch observations were made of ECHA J0841.5--7853 and
ECHA J0843.3--7905 during 1999 February--March and 2000
February--March as part of a larger study to measure the rotation
periods of the late-type RECX stars in the $\eta$ Cha cluster.
Lawson et al. (2001) found that all of these stars were
variable, with periods attributed to rotational modulation
of cool starspots.  Differential $V$-band observations of
the 2 new cluster members were obtained with respect to
nearby stars within their CCD frames.  The 1999 and 2000
data sets for each star were independently analysed using
the Lomb-Scargle Fourier method for non-equally spaced data,
with the frequency range of $f = 0-2$ d$^{-1}$ examined for
periodicities.  Both stars showed periodic variations in each
year.  Phased light curves are shown in Fig. 5, and details
of the periodicities are listed in Table 2.  The S/N ratio
for each periodicity was determined by measuring the residual
noise level in the pre-whitened data sets.

For ECHA J0841.5--7853, a low-amplitude 1.73-d
periodicity was recovered in each year.  As with the late-type
RECX stars, we associate the variations with the rotational
modulation of cool spots.  Following Allain et al. (1996) we
calculated the lower limit to the fractional spot coverage by
assuming the spots were dark.  The spot fractions are listed
as percentages in Table 2.  Lawson et al. (2001) corrected
the {\it VRI\,} photometry of the RECX stars for the effects
of the starspots by estimating the $V$ mag and colours of
the unspotted, or minimum-spotted, star from the phase of
the {\it VRI\,} photometry and the $V$-band amplitude of the
star.  The {\it VRI\,} measurements listed in Table 1 were
obtained near maximum light and thus represent the unspotted,
or minimum-spotted, star.

For ECHA J0843.3--7905, a 12-d periodicity was measured in each
year, but with different structure in the light curve.  During
1999 February, a short-duration ($2-3$ d) peak was observed
that may have been an optical flare.  Underlying the peak of
$\Delta V = 0.65$ mag is a $0.2-0.3$ mag amplitude
quasi-sinusoidal variation that dominates the Fourier analysis.
The light curve during 2000 has a different appearance, but
with the same underlying period and $V$-band amplitude.  The
{\it VRI\,} photometry reported in Table 1 was obtained in
2000 near maximum light.  Given the large
photometric amplitude and the variable structure of the
light curve probably driven by accretion hotspots, these
data may not represent the unspotted photospheric values.

\begin{table}
\centering
\caption{Periods present in the $V$-band light curves during
1999 and 2000.  The final column gives lower limits to the
dark spot coverage for ECHA J0841.5-7853.  ECHA J0843.3-7905
is assumed to have variability due to bright features; see
Section 3.2 for details.}
\begin{tabular}{@{}cccccc@{}}
\hline
ECHA          & Year & Period & Amp.   & S/N   & Coverage \\
               &      &   (d)  &  (mag) & ratio & (per cent) \\ \hline
J0841.5--7853 & 1999 & 1.73   & 0.04   & 3     & 4 \\
               & 2000 & 1.73   & 0.07   & 3     & 6 \\
J0843.3--7905 & 1999 & 12.8   & 0.25   & 4     & -- \\
               & 2000 & 12.2   & 0.44   & 5     & -- \\
\hline
\end{tabular}
\end{table}

\begin{figure}
\begin{center}
\epsfxsize=8.4cm
\epsffile{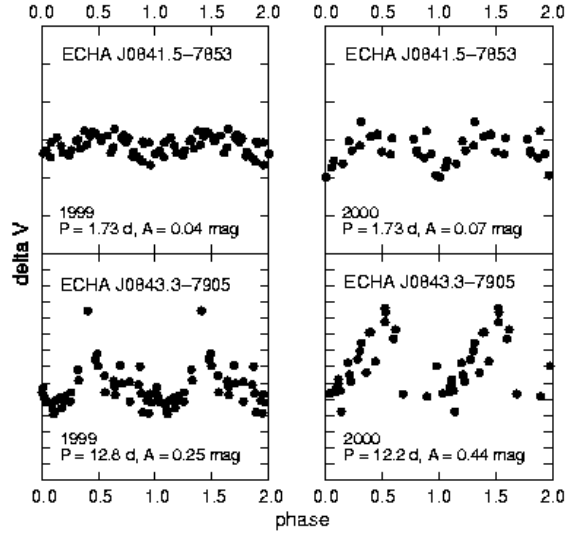}
\caption{Phase-folded $V$ light curves for ECHA J0841.5--7853 and
ECHA J0843.3--7905 obtained during 1999 (left panels) and 2000
(right panels). Within each panel, the period and peak-to-peak
amplitude determined from the Fourier analysis of the light curves
is shown.  Ordinate tick marks are separated by 0.1 mag in all
cases.  These results are discussed in Section 3.2.}
\end{center}
\end{figure}

\subsection{Comments on the rejected candidates}

Analysis of the MSSSO spectra showed 3 of these stars (GSC 9402\_1003,
GSC 9398\_0099 and GSC 9403\_0831) were late-K or early-M spectral
type giants with H$\alpha$ in narrow absorption with {\it EW\,}
$\approx 1$ \AA\, and no detectable $\lambda 6707$ Li I absorption
(see Table 1).  All 3 stars can also be ruled out as cluster members
from their proper motions.  GSC 9398\_0099 and GSC 9403\_0831 have
Tycho-2
(H$\o$g et al. 2000) proper motions ($\mu_{\alpha}$, $\mu_{\delta}$) =
(3.7, 1.4) and ($-9.5$, 6.2) mas\,yr$^{-1}$, respectively (1$\sigma$
uncertainties are $3-4$ mas\,yr$^{-1}$).  GSC 9402\_1003 has a USNO
UCAC1 (Zacharias et al. 2000) proper motion of ($\mu_{\alpha}$,
$\mu_{\delta}$) = ($-0.3$, 25.3) mas\,yr$^{-1}$ (1$\sigma$, 12.5
mas\,yr$^{-1}$).  These values differ by $> 2 \sigma$ from the
{\it Hipparcos\,}/Tycho-2 proper motion for the cluster of
($\mu_{\alpha}$, $\mu_{\delta}$) = ($-30.0 \pm 0.3$, $27.8 \pm 0.3$)
mas\,yr$^{-1}$ (Mamajek et al. 2000)\footnote{Note that Table 2 of
Mamajek et al.  (2000) incorrectly lists $\mu_{\delta} = -27.8$
mas\,yr$^{-1}$.}.

The remaining candidate is listed as `Anonymous' in Table 1, since
we found no catalogue entries for this star.  Both the spectrum and
the colours of the star are consistent with a M5 spectral classification.
The star has weak H$\alpha$ emission and no detectable $\lambda 6707$
Li I absorption line.  Study of on-line scanned plates used to compile
the USNO-A2.0 catalogue (Monet et al. 1998) show the star has high
proper motion (see Fig. 6).  We determined the position of the star
against USNO-A2.0 positions for several nearby stars at 4 epochs
(1978.10, 1986.18, 1996.13 and 2000.12); the first 3 from scanned
plates available from USNO, and the last from analysis of a SAAO
CCD image.  From these positions we derived
($\mu_{\alpha}$, $\mu_{\delta}$) $\approx$ ($-140$, 360) mas\,yr$^{-1}$.
The USNO-A2.0 catalogue rejects stars with $\mu > 300$ mas\,yr$^{-1}$.
The star is likely a dMe.  If it is a main-sequence star, then the
star has a distance of $\sim 50$ pc and tangential velocity of
$\sim 100$ km\,s$^{-1}$.

\begin{figure}
\begin{center}
\epsfxsize=8.0cm
\epsffile{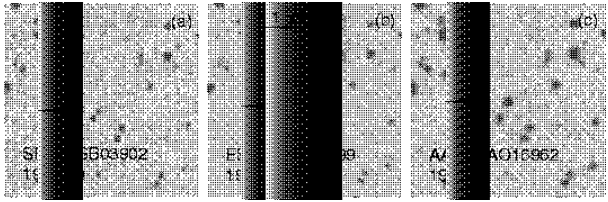}
\caption{USNO finder charts (width = 2.5 arcmin) for the
`Anonymous' candidate, obtained from Schmidt plates
taken at epochs (a) 1978.10, (b) 1986.18 and (c) 1996.13.  The
plate source (SRC = Science and Engineering Research Council,
ESO = European Southern Observatory, AAO = Anglo-Australian
Observatory), emulsion ($J$ or $R$) and plate number is given.}
\end{center}
\end{figure}

\section{Summary and Conclusions}

Hillenbrand \& Meyer (1999) examined the frequency of disks as
a function of stellar age for nearby clusters and star forming
regions and found that disks disperse on a timescale of $< 10$
Myr, with few disks remaining at ages $> 15$ Myr.  Haisch, Lada
\& Lada (2001) conducted a similar study on the stellar population
of several young clusters, concluding that essentially all stars
lose their disks within $\sim 6$ Myr.  The prevalence of disks
in the $\approx 9$ Myr-old $\eta$ Cha cluster will be examined
by Lyo et al. (in preparation).

An `old' and nearby CTT star such as ECHA J0843.3--7905 is
therefore rare.  A similar star is the CTT star TW Hya, also
nearby ($d = 56$ pc) and $\approx 10$ Myr old (Webb et al. 1999).
The resolved pole-on disk surrounding TW Hya has been a focus
for study of disk structure (e.g. Trilling et al. 2001) and
the early planet formation environment.  ECHA J0843.3--7905
is of later spectral type and lower mass than TW Hya.
Imaging studies of ECHA J0843.3--7905 could give valuable
insight into the nature of evolved disks (and indirectly
planets) around dwarf M stars.

Like other CTT stars, ECHA J0843.3--7905 has a strong infrared
excess.  Analysis of $L$-band imaging of the cluster core
obtained with the South Pole Infrared Explorer telescope during
1999 (Lyo et al., in preparation) found $L = 7.8$ for this star.
Assuming an M2 spectral type, the $L$-band excess is $\approx 2$
mag.  The {\it IRAS\,} FSC entry for the star indicates high-quality
25- and 60-$\mu$m fluxes.

Our survey of $\approx 40$\% of the known extent of the cluster
found 2 new cluster members not detected by the discovery
{\it ROSAT\,} HRI image of Mamajek et al. (1999, 2000), thereby
increasing the number of stellar primaries to 15.  A survey of
similar depth across the cluster might therefore find only
several more new members.  This result
appears to be at odds with Mamajek et al. (2000) who predicted
from consideration of the cluster IMF that the stellar population
was $2-4 \times$ the (then) known number of 13 primaries.  Our
survey might indicate that the cluster extent is not constrained
by the {\it ROSAT\,} HRI field and that $15-40$ primaries await
discovery beyond the HRI boundary.  Alternatively, if the cluster
is constrained by the HRI field, then the low success
rate of our study suggests the cluster may contain as few as
$\approx 20$ primaries.  (These estimates do not address the
brown dwarf population expected to accompany the stellar members.)
Either of the above population scenarios indicates the X-ray
survey must have been relatively complete at detecting cluster
members within the HRI field.  This result confirms the unusual
skewness of the
ratio of X-ray luminosity to bolometric luminosity noted by
Mamajek et al.  (1999, 2000), with most of the late-type RECX
stars having a flux ratio near the `saturation' level of
log $L_{X}$/$L_{\rm bol} \approx -3$.

\section*{Acknowledgments}

We thank the SAAO and MSSSO time allocation committees for
telescope time during 1999 and 2000.  WAL and LAC thank the
staff of SAAO for their assistance.  WAL and EEM thank the
MSSSO support staff for their guidance in operating the 2.3-m
telescope.  WAL acknowledges financial support from the
Australian Research Council Small Grant Scheme and University
College Special Research Grants.  LAC is supported by a NRF
Post-graduate Scholarship.  EEM thanks the SIRTF Legacy Science
Program for support.  EDF's research is supported in-part by
NASA contracts NAS8-38252 and NAG5-8422.  This research made
use of the POSS-II survey and the VizieR database.  The
scanned plates used to construct the USNO-A2.0 catalogue
are available at {\tt http://www.nofs.navy.mil}.

\bsp

\label{lastpage}

\end{document}